\documentclass[paper]{JHEP}
\usepackage{amsmath}
\usepackage{epsfig}

\def\as{\alpha_{\mbox{\scriptsize s}}}
\def\bas{{\bar\alpha_{\mbox{\scriptsize s}}}}

\def\ee{e^+e^-}

\def\NP{\mbox{\scriptsize NP}}

\def\eps{\epsilon}
\def\de{\delta}
\def\De{\Delta}
\def\Om{\Omega}
\def\om{\omega}
\def\gam{\gamma}
\def\lam{\lambda}
\def\cp{\lambda^{\NP}}

\def\Tin{\Theta_{\mbox{\scriptsize in}}}
\def\Cin{{\cal C}_{\mbox{\scriptsize in}}}
\def\thin{{\theta}_{\mbox{\scriptsize in}}}

\def\RR{{\rm RR}}

\def\thc{\theta_{\rm crit}}
\def\Eout{E_{\mbox{\scriptsize out}}}
\def\Tout{\Theta_{\mbox{\scriptsize out}}}
\def\Cout{{\cal C}_{\mbox{\scriptsize out}}}
\def\Qout{{E}_{\mbox{\scriptsize out}}}
\def\xB{x_{\rm B}}

\def\cO#1{{\cal{O}}\left(#1\right)}
\def\half{\mbox{\small $\frac{1}{2}$}}

\def\VEV#1{\left\langle#1\right\rangle}

 \newskip\humongous \humongous=0pt plus 1000pt minus 1000pt
   \newif\ifdtup

\def\la{\mathrel{\mathpalette\fun <}}
\def\ga{\mathrel{\mathpalette\fun >}}
\def\fun#1#2{\lower3.6pt\vbox{\baselineskip0pt\lineskip.9pt
  \ialign{$\mathsurround=0pt#1\hfil##\hfil$\crcr#2\crcr\sim\crcr}}}

\title{Away-from-jet energy flow\footnote{Research supported in part
    by the EU Fourth Framework Programme, `Training and Mobility of
    Researchers', Network `Quantum Chromodynamics and the Deep
    Structure of Elementary Particles', contract FMRX-CT98-0194 (DG12
    - MIHT).}}

\author{
A.~Banfi, G.~Marchesini, G.~Smye\\
Dipartimento di Fisica, Universit{\`a} di Milano--Bicocca and \\
INFN, Sezione di Milano, Italy
}

\abstract{We consider interjet observables in hard QCD processes given
  by the energy flow $\Eout$ in a region away from all hard jets. Here
  the QCD radiation is depleted ($\Eout\ll Q$), and therefore these
  observables provide ideal means to study non-perturbative effects.
  We derive an evolution equation (in the large $N_c$ limit) which
  resums, for large $Q/\Eout$, all leading terms arising from large
  angle soft emission (double logarithms are absent). We discuss the
  analytical features of the result and identify universal and
  geometry-dependent contributions. Our analysis confirms features
  found using numerical methods by Dasgupta and Salam.}


\preprint{
     Bicocca--FT--02/10\\
     hep-ph/0206076\\
     June 2002}
\begin{document}

\section{Introduction}
Study of ``interjet'' hadronic emission \cite{MW,DS,BKS} is of special
interest in QCD. In general, such emission originates from the flow of
colour between jets, and therefore its analysis is important for
understanding the mechanism of the overall colour neutralization. In
practical terms, the interjet radiation is typically soft and so the
distributions are sensitive to possible non-perturbative components
such as underlying events. For recent and not-so-recent experimental
studies in hadronic colliders, see \cite{UA1,FHT}.

A characteristic feature of inclusive interjet distributions is that
collinear singularities do not play a relevant r\^ole and that the
distributions are essentially determined by large angle soft
emissions, which are responsible for the coherence of QCD radiation
\cite{coherence,BCM}.

An example of an interjet quantity is $\Eout$, the total energy (or
transverse momentum) of hadrons emitted in a region $\Cout$ away from
all hard jets.  The $\Eout$ distribution is infrared (and collinear)
safe so that all its perturbative (PT) coefficients are finite and
computable (in principle, although the resulting series then does not
converge). Typically one has $\Eout\ll Q$, with $Q$ the hard scale of
the process, so that reliable QCD estimates require the
resummation\footnote{Only the average value of $\Qout$ or the
  distribution for relatively large $\Eout$ are obtained by finite
  order calculations \cite{MW}.} of the logarithmically enhanced terms
generated by soft emitted or virtual partons. Since for this
observable no logarithms are generated from collinear singularities,
one has that the leading contributions to the distribution are given
by single logarithmic (SL) terms $\as^n\ln^n Q/\Eout$.

This should be contrasted with the case of observables in which the
measured region includes one or more jets. An example of this is the
thrust in the $\ee$ process. In such cases both soft and collinear
singularities contribute, so that the logarithm of the distribution
contains, besides SL terms, also double logarithmic (DL) terms ($\as^n
L^{n+1}$ with $L$ a large logarithm).

Typically one expects that less singular distributions would be more
difficult to resum. This is actually the case here.  For a
distribution in a ``global'' observable, i.e. involving the entire
phase space, the PT expansion can be resummed \cite{PTglobal} by means
of a {\it linear} evolution equation (a generalization of the DGLAP
equation \cite{DGLAP}) based on the factorization of collinear
singularities and coherence of the QCD branching structure
\cite{coherence,BCM}. The resulting distribution can be expressed as a
Sudakov form factor (the exponential of a {\it radiator}) which
corresponds to bremsstrahlung emission directly from the primary
partons.

As shown by Dasgupta and Salam \cite{DS2}, ``non-global''
observables\footnote{Interjet observables are a particular case of
  non-global observable. Their numerical study for $\ee$ has been
  performed in \cite{DS}.}, i.e. involving a part of the phase space,
cannot be described only by the bremsstrahlung process but they
involve the entire structure of successive parton branching.
Calculations in $\ee$ and DIS were performed numerically by a Monte
Carlo method based on dipole branching emission derived from the
distribution of many soft gluons emitted off a colour singlet dipole
of hard partons given in \cite{BCM,FMR} in the large $N_c$ limit.

Given the relevance of interjet observables, it would be important to
have an analytic formulation for the SL resummation and this is the
aim of the present study.  We derive an evolution equation based on
the soft multi-gluon distribution given in \cite{BCM,FMR} valid for
soft emissions in all angular regions.  This evolution equation resums
all SL contributions for the interjet distribution. It is based on
energy ordering but implies also angular ordering
\cite{coherence,BCM}, which is the basis of Monte Carlo QCD
simulations \cite{MC}.  As in \cite{DS} this formulation of the
branching involves colour singlet dipoles in the large $N_c$
approximation.

In the main text we consider the case of $\ee$ annihilation with an
unobserved region $\Cin$ defined by a cone around the thrust axis. Our
analysis is extended in Appendix \ref{App:DIShh} to processes with
incoming hadrons such as DIS and hadron-hadron collisions with large
$P_t$-jets. We show that for all these hard processes the interjet
distributions are given (in the large $N_c$ limit) in terms of a
single function which depends on the geometry of the process.

We derive the shape of the interjet distribution for large $Q/\Eout$
and the structure of the branching and we discuss emission into the
$\Cout$ region. We characterize these aspects in terms of quantities
which are either universal or geometry-dependent. We confirm features
found in \cite{DS} in numerical studies.

The paper is organized as follows.  In section \ref{sec:Observable} we
describe the observable we consider and in section \ref{sec:QCD} we
express the interjet distribution in terms of soft multi-parton
emission.  In section \ref{sec:evev} we derive the evolution equation
and we compare the case of the interjet distribution with the ones of
other observables.  In section \ref{sec:gen} we describe the general
features of the distribution, with both analytical and numerical
methods.  Section \ref{sec:Large} contains the detailed technical
discussion for large $Q/\Eout$.  Universal and geometry-dependent
features are derived. Finally, in section \ref{sec:disc} we recall the
main physical features of the interjet distribution.

\section{The observable \label{sec:Observable}}
For a given hard process we consider as our interjet observable the
total energy of hadrons emitted in a phase space region $\Cout$ away
from all hard jets
\begin{equation}
  \label{eq:Eout}
  \Eout=\sum_{h\,\in\,\Cout}\om_h\>.
\end{equation}
Most of the energy flows inside the jet regions so that typically we
have $\Eout\ll Q$ with $Q$ the hard scale. The features we will
describe for this observable can be extended to other interjet
observables such as the total hadron transverse energy.  The interjet
region $\Cout$ could be defined in various ways according to the hard
process.  The complementary region including the jets will be denoted
by $\Cin$.

In $\ee$ annihilation we will consider the interjet region $\Cout$ as
\begin{equation}
  \label{eq:region-ee}
  -\cos\thin<\cos \theta_h<\cos\thin\>,
\end{equation}
with $\theta_h$ the angle with respect to the thrust axis $\vec{n}_T$.
We will discuss the following collinear and infrared safe distribution
\begin{equation}
  \label{eq:Sig-ee}
  \Sigma_{\ee}(\Qout)=\sum_n\int\frac{d\sigma_n}{\sigma_T}\cdot
  \Theta\!\left(\Qout-\sum_{h\,\in\,\Cout}\om_h\right),
\end{equation}
with $d\sigma_n$ the $n$-hadron distribution and $\sigma_T$ the total
cross section. The dependence on $\thin$ and $Q$ is understood. This
distribution is normalized to one at the kinematical limit $\Eout\sim
Q$.  The process is dominated by two-jet events aligned along the
thrust axis so that $\Eout$ is typically much smaller then $Q$.
Three-jet events, with $\Eout\sim Q$, are of order of $\as(Q)$.

In Appendix \ref{App:DIShh} we discuss examples of a similar interjet
observable in DIS and hadron-hadron collisions with high $P_t$-jets.

\section{QCD resummation \label{sec:QCD}}
At parton level, for small $\Eout$ the distribution in $\ee$ is
described by the emission of the primary quark-antiquark pair $p\bar
p$ accompanied by secondary soft gluons $k_i$,
\begin{equation}
  \label{eq:proc}
  \ee\to p\, k_1\ldots k_n \,\bar p\>.
\end{equation}
In the soft limit, the primary quark and antiquark are aligned along
the thrust axis. The leading contribution of the soft multi-parton
distribution $M^2_n$ is obtained by considering strong energy ordering
for the emitted partons. Here, the phase space integration can be
approximated by ($E\sim Q$)
\begin{equation}
  \label{eq:dPhin}
  d\Phi_n=\prod_i\om_i d\om_i\frac{d^2 \Om_i}{4\pi}\Theta(E-\om_i)\>.
\end{equation}
For any one of the $n!$ strongly energy-ordered regions, the real
emission contribution to the soft multi-parton distribution is given,
in the large $N_c$  limit, by the factorized expression (see \cite{BCM,FMR})
\begin{equation}
  \label{eq:M2n}
  \begin{split}
    &M^2_n(p k_1 \ldots k_n \bar p)=M_0^2(p\bar p)\cdot 
    S_{p\bar p}(k_1\ldots k_n)\>, 
  \\
    &S_{p\bar p}(k_1\ldots k_n)=
    \frac{1}{n!}\,\prod_i\frac{\bas}{\om_i^2}\,
    \sum_{\pi_n} W_n(p k_{i_1} \ldots k_{i_n} \bar p)\>,
    \qquad \bas=N_c\frac{\as}{\pi}\>,
  \end{split}
\end{equation}
where the sum is over all $n!$ permutations. For the fundamental
permutation we have
\begin{equation}
  \label{eq:Wn}
  W_n(p k_{1} \ldots k_{n} \bar p)\>=\>\frac{(p\bar p)}
  {(pk_1)(k_1k_2)\ldots(k_n\bar p)}\>,\qquad
  (qq')\equiv 1-\cos\theta_{qq'}\>.
\end{equation}
This distribution has both soft ($\om_i\to0$) and collinear
($\theta_{qq'}\to0$) singularities. It is valid in any strongly
energy-ordered region. No collinear approximations are involved in the
derivation, so it is valid even at large angles as needed for our
study. 

To leading order for small $\Eout$, the distribution
$\Sigma_{\ee}(\Eout)$ is obtained by using
\begin{equation}
  \label{eq:dsigman}
  \frac{d\sigma_n}{\sigma_T}=
  S_{p\bar p}(k_1\ldots k_n)\cdot d\Phi_n\>.
\end{equation} 
Here we have to add to the soft distribution given in \eqref{eq:M2n}
the virtual corrections to the same order in the soft limit.  This
will be done in the next section in which we derive the evolution
equation giving $\Sigma_{\ee}(\Eout)$.  The expression
\eqref{eq:dsigman}, being accurate at the leading order in the soft
limit, contains all the SL terms we want to resum.

The basis for the resummation is the factorized structure of
$M^2_{n}$. We then factorize\footnote{Alternatively, we can use strong
  energy ordering so that $\Eout$ is the energy of the hardest
  secondary soft gluon.  The Mellin transform method simplifies the
  combinatorics.}  also the theta function in \eqref{eq:Sig-ee} by
writing
\begin{equation}
  \label{eq:Theta-fac}
  \Theta\left(\!\Eout\!-\!\sum_{i\in\Cout}\om_i\!\right)=
  \int\frac{d\nu\,e^{\nu \Eout}}{2\pi i \nu}\, \prod_i u(k_i)\>,
  \quad   u(k)=\Tin(k)+e^{-\nu\om}\Tout(k)\>,
\end{equation}
where $\Tout(k)$ and $\Tin(k)$ are the support functions in the
interjet region $\Cout$ and $\Cin$ respectively. The geometry of the
interjet region $\Cout$, i.e. the $\thin$ dependence, is totally
contained in the source $u(k)$.  The distribution is then given by
\begin{equation}
  \label{eq:Sig-ee1}
  \Sigma_{\ee}(\Eout)\>=\>\int\frac{d\nu\,e^{\nu\Eout}}{2\pi i \nu}
  \>G_{p\bar p}(E,\nu^{-1})
\simeq G_{p\bar p}(E,\Eout)\>,
\end{equation}
where the Mellin variable $\nu$ runs along the imaginary axis to the
right of the singularities of $G_{p\bar p}(E,\nu^{-1})$.  In
\eqref{eq:Sig-ee1} we have evaluated
the Mellin integration by steepest descent to give $\nu\simeq\Eout^{-1}$.
The contribution from real emission \eqref{eq:M2n} is given by
\begin{equation}
  \label{eq:GR-ee}
  G_{p\bar p}^{({\rm real})}(E,\Eout)=1\!+\!
\sum_{n=1}^{\infty}\!\int\prod_{i=1}^n
\left\{\bas \frac{d\om_i}{\om_i}\frac{d^2 \Om_i}{4\pi}u(k_i)\,
\Theta(E\!-\!\om_i)\!\right\} W_n(p k_{1} \ldots k_{n} \bar p)\,.
\end{equation}
Here we set $\nu=\Eout^{-1}$ in the source $u(k)$ and used the
symmetry in the secondary gluons to select the fundamental permutation
at the cost of the $1/n!$ factor.  Virtual corrections to
\eqref{eq:GR-ee} will be introduced in the next section.

In Appendix \ref{App:DIShh} we extend this analysis to hard processes
with incoming hadrons and show that interjet distributions for all
hard processes are described by the same function
$G_{p_ip_j}(E,\Eout)$ with $p_i$ and $p_j$ the direction of pairs of
hard jets (incoming or outgoing) and on the interjet region $\Cout$.

\section{Evolution equation \label{sec:evev}}
To formulate an evolution equation we need to introduce the
distribution $G_{ab}$ for a general pair of primary partons $p_ap_b$,
obtained from \eqref{eq:GR-ee} by replacing $p\bar p$ by generic
dipole momenta $p_ap_b$.  This distribution depends on the direction
of $p_ap_b$ with respect to the $\ee$ thrust axis and on the geometry
of the interjet region $\Cout$, i.e. on $\thin$ in
\eqref{eq:region-ee}. To obtain the evolution equation we use
\begin{equation}
  \begin{split}  
    &E\partial_E\left\{W(p_ak_1\ldots k_n p_b)
    \prod_{i=1}^n\Theta(E\!-\!\om_i)\right\}
    =\sum_{\ell=1}^nE\de(E\!-\!\om_{\ell})\,w_{ab}(k_{\ell})
  \\ 
    &\cdot\left\{W(p_ak_1\ldots k_{\ell})
    \prod_{i=1}^{\ell-1}\Theta(E\!-\!\om_i)\right\} \cdot
    \left\{W(k_{\ell}\ldots k_n p_b)
    \prod_{i=\ell+1}^{n}\Theta(E\!-\!\om_i)\right\} \>,
  \end{split}
\end{equation}
where 
\begin{equation}
  \label{eq:wab}
  w_{ab}(k)=\frac{(p_ap_b)}{(p_ak)(kp_b)}=
  \frac{1-\cos\theta_{ab}}{(1-\cos\theta_{ak})(1-\cos\theta_{kb})}\>.
\end{equation}
We then deduce the basic equation ($\nu=\Eout^{-1}$ dependence is
understood)
\begin{equation}
  \label{eq:basic}
\begin{split}
  E\partial_E\>G_{ab}(E)=\int \frac{d^2 \Om_k}{4\pi}\,\bas
  w_{ab}(k)\left[u(k)\,G_{ak}(E)\cdot G_{kb}(E)-G_{ab}(E)\right].
\end{split}
\end{equation}
Here we have added virtual corrections, the last term in the square
bracket, to the same order as the real emission contributions, see
\cite{BCM}.  This evolution equation corresponds to soft dipole
emission with energy ordering. Since $w_{ab}(k)$ effectively
constrains $k$ into the angular region within the $ab$ dipole,
\eqref{eq:basic} also implies angular ordering (after azimuthal
averaging).  Large angle regions are correctly taken into account.

It is convenient to write \eqref{eq:basic} in the form
\begin{equation}
  \label{eq:basic1}
  \begin{split}
  E\partial_E\>G_{ab}(E)
  &=-E\partial_E\>R_{ab}^{(0)}(E)\cdot G_{ab}(E)\\
  &+\int \frac{d^2 \Om_k}{4\pi}\,\bas
  w_{ab}(k)\,u(k)\,\left[\,G_{ak}(E)\cdot G_{kb}(E)-G_{ab}(E)\,\right],
  \end{split}
\end{equation}
with $R_{ab}^{(0)}(E)$ the SL Sudakov radiator for the bremsstrahlung
emission 
\begin{equation}
  \label{eq:Rad}
  R_{ab}^{(0)}(E)\>=\>\int_0^E\frac{d\om}{\om}
  \int\frac{d^2 \Om_k}{4\pi}\,\bas\,w_{ab}(k)\,
  [1-u(k)]\>=\>\De\cdot r_{ab}\>,
\end{equation}
where $\De$ depends on $E,\Eout$ and $r_{ab}$ on the geometry of
the interjet region \eqref{eq:region-ee}
\begin{equation}
  \label{eq:rad}
  \De=\int_{0}^{E}\frac{d\om}{\om}\,\bas\,
  \left[ 1-e^{-{\om}/{\Eout}} \right]\>, \qquad
  r_{ab}=\int_{\Cout}\frac{d^2 \Om_k}{4\pi}\,w_{ab}(k)\>.
\end{equation}
Here we have used $[1\!-\!u(k)]\!=\![1\!-\!e^{-\om/\Eout}]\,\Tout(k)$
which entails that, in the unobserved jet region $\Cin$, the infrared
and collinear singularities of $w_{ab}(k)$ are fully cancelled between
real and virtual contributions.

In the quantity $\De$ the argument of the running coupling $\as$ is
determined by the successive hard emission contributions and cannot be
determined in the present analysis. Detailed analysis \cite{CMW} shows
that, in the physical scheme, it is given by the transverse momentum
relative to the $ab$-dipole which here can be approximated by the
energy since the contributions to this radiator come from large angle
emission.

For small $\Eout$ we can evaluate $\De$ by the standard SL
approximation
\begin{equation}
  \label{eq:SL}
1-e^{-\nu \om}\,\simeq\, \Theta(\om-e^{\gam_E}\nu^{-1})
\,\simeq\, \Theta(\om-\Eout)\,,
\qquad \De\,\simeq\,\int_{\Eout}^{E}\frac{d\om}{\om}\,\bas\>.
\end{equation}
The radiator $R_{ab}^{(0)}$ is then the contribution from the virtual
parton in the $\Cout$ phase space with energy $\om \ga \Eout$.  For
fixed $\as$ we have $\De\simeq\bas\ln E/\Eout$.

The evolution equation \eqref{eq:basic1} does not generate any
collinear logarithms, either from the bremsstrahlung factor $r_{ab}$
(since the region $\Cout$ does not include $p_a$ or $p_b$) or from the
integral term in the second line of \eqref{eq:basic1}. The latter is
collinear regular due to cancellation between the branching
$G_{ak}\cdot G_{kb}$ and the virtual $-G_{ab}$ terms.  Indeed, for $k$
collinear to $p_a$ one has $G_{ak}\to1$ and $G_{kb}\to G_{ab}$ so that
the sum in the square brackets vanishes and regularizes the
singularity of $w_{ab}(k)$. Since only soft singularities remain, the
leading terms of $G_{ab}$ are SL contributions.

To SL order, we can replace $u(k)$ by $\Tin(k)$ in the integral term
of \eqref{eq:basic1}. To show this observe that the remaining piece
$e^{-\om/\Eout}\,\Tout(k)$ of the source contributes with energy
$\om\la\Eout$ (see \eqref{eq:SL}) while the softest gluon, which
contributes to $R_{ab}^{(0)}$, has a typically larger energy
($\om\ga\Eout$). We conclude that the soft secondary branching takes
place in $\Cin$ and only the final parton enters the interjet region
$\Cout$ and here it contributes only with the virtual correction with
energy $\om\ga\Eout$.

By replacing $u(k)\to \Tin(k)$ in the branching term of
\eqref{eq:basic1} we have that the distribution $G_{ab}$ depends on
$E$ and $\Eout$ only through the function $\De$. We then can replace
$G_{ab}(E,\Eout)\to G_{ab}(\De)$ and obtain, to SL accuracy,
\begin{equation}
  \label{eq:basic2}
  \partial_{\De}\,G_{ab}(\De)\!=\!-r_{ab}\,G_{ab}(\De)
  +\!\int_{\Cin}\! \frac{d^2 \Om_k}{4\pi}\,
  w_{ab}(k)\left[\,G_{ak}(\De)\cdot G_{kb}(\De)-G_{ab}(\De)\,\right],
\end{equation}
with the initial condition $G_{ab}(\De\!=\!0)\!=\!1$.  The
distribution $G_{ab}(\De)$ can be factorized into two pieces
\begin{equation}
  \label{eq:gab}
  G_{ab}(\De)=e^{-R_{ab}^{(0)}(\De)}\cdot g_{ab}(\De)\>.
\end{equation}
The first is the Sudakov factor given by bremsstrahlung emission from
the primary hard partons $p_ap_b$. The second factor is the result of
successive soft secondary branching which satisfies
\begin{equation}
  \label{eq:basic3}
  \partial_{\De}\,g_{ab}(\De)\!=\!
  \!\int_{\Cin}\! \frac{d^2 \Om_k}{4\pi}\,
  w_{ab}(k)\left[\,U^{(0)}_{abk}(\De)\,
  g_{ak}(\De)\cdot g_{kb}(\De)-g_{ab}(\De)\,\right],
\end{equation}
where
\begin{equation}
  \label{eq:U0}
  U^{(0)}_{abk}(\De)= 
  e^{-R_{ak}^{(0)}(\De)-R_{kb}^{(0)}(\De)+R_{ab}^{(0)}(\De)}\>,
\end{equation}
is an effective source. Notice that \eqref{eq:basic3} has the same
structure as \eqref{eq:basic} with $u$ replaced by $U^{(0)}$.
The factorization in \eqref{eq:gab} can then be iterated, see later.

\subsection{Comparison with global observables \label{sec:global}}
The evolution equation \eqref{eq:basic1} can be generalized to any
collinear and infrared safe observable. Consider
for example an additive global observable
\begin{equation}
  \label{eq:V} V=\sum_i v(k_i)\>,
\end{equation}
with $v(k_i)$ a function of the emitted hadron momentum. One
deduces\footnote{In general one may need to factorize phase space
  constraints also, see \cite{PTglobal}.} again \eqref{eq:basic1} with
the bremsstrahlung radiator involving the source for the specific
observable
\begin{equation}
  \label{eq:Rad-gen}
  R_{ab}^{(0)}(E)\>=\>\int_0^E\frac{d\om}{\om}
  \int\frac{d^2 \Om_k}{4\pi}\,\bas\,w_{ab}(k)\,[1-e^{-\nu v(k)}]\>.
\end{equation}
This quantity is collinear and infrared safe if $v(k_i)$ vanishes for
$k_{ti}\to0$.  In this case the radiator involves also double
logarithmic contributions $\as^n\ln^{n+1}V$ which arise from the
collinear and infrared singular structure of $w_{ab}(k)$.  SL terms
coming from hard collinear emission are not included in
\eqref{eq:Rad-gen}, but these can easily be included by adding to
$w_{ab}(k)$ the finite pieces of the splitting functions.

The soft secondary branching in the evolution equation (second line in
\eqref{eq:basic1}) plays a very different r\^ole\footnote{As recalled
  before, the present treatment does not include hard contributions
  from the secondary branching which reconstruct the running coupling
  argument.} for global and local observables.  As previously
discussed, for local observables all soft secondary branching terms
contribute to SL accuracy.  For global observables, instead, soft
secondary branching contributes beyond SL accuracy \cite{coherence,BCM}.
This fact can be easily seen in the present formulation.  

Consider for instance the two loop contribution which involves the
combination of the sources $u(k_2)[1-u(k_1)]$ with $\om_1<\om_2$.
For the local observable $\Eout$ here considered, we have that
$u(k_2)[1-u(k_1)]$ is consistent with energy ordering provided $k_1\in
\Cout$ and $k_2\in \Cin$ so that one has a SL contribution
$\as^2\ln^2E/\Eout$.
For a global observable instead, with $\Cin=0$, we have that the
combination $u(k_2)[1-u(k_1)]$ contributes in the region
$v(k_2)>v(k_1)$ (see \eqref{eq:SL}) which is in conflict with energy
ordering. As a result, secondary branching has no infrared logarithms
and one remains with a subleading term, $\as^2\ln V$, which comes from
the collinear singularity of the dipole emission. As a result for a
global observable, the distribution is simply given by a Sudakov form
factor, the exponentiation of the DL bremsstrahlung radiator.

In \eqref{eq:Sig-ee} we have considered interjet observables in which
$\Cout$ does not include the primary parton region. The analysis can
be generalized to the case in which the phase space region defining
the local observable includes one primary parton \cite{DS2}.  Here one has DL
contributions and soft secondary branching giving SL
terms to all orders.

\section{General features of the distribution \label{sec:gen}}
We cannot solve analytically the SL evolution equation
\eqref{eq:basic2}. However we can study various aspects of the
distribution using different approaches to give a picture of how the
distribution behaves. We first observe that the distribution satisfies
the constraint
\begin{equation}
  \label{eq:unity}
  0\le G_{ab}(\De)\le 1\>,
\end{equation}
for any $\De$ and $ab$. A proof of this is given in Appendix
\ref{App:Proof}. We also know that $G_{ab}(0)=1$ and that, since the
$ab$-dipole does not emit for $a\!=\!b$ we have
\begin{equation}
  \label{eq:constraint}
  G_{ab}(\De)\>\to\>1\>,\qquad \mbox{for}\quad a\to b\>,
\end{equation}
for any $\De$.  For $a\!\ne\! b$, $G_{ab}(\De)\to 0$ as $\De$ becomes
large.  This implies that the distribution $G_{ab}(\De)$ has a peak at
$a\!=\!b$, whose width decreases with increasing $\De$. This peak
governs the evolution of the entire distribution.

We have used three approaches to study the distribution: an iterative
solution valid at small $\De$, the asymptotic behaviour at large
$\De$, and numerical solution. We discuss the results in the following
subsections.

\subsection{Iterative solution at small $\De$ \label{sec:iterative}}
We observe that \eqref{eq:basic3} for $g_{ab}$ is similar to
\eqref{eq:basic} for $G_{ab}$ with $u(k)$ replaced by $U^{(0)}_{abk}$
defined in \eqref{eq:U0}. By iterating the procedure used to factorize
the bremsstrahlung piece (see \eqref{eq:gab}) we obtain the general
expansion
\begin{equation}
  \label{eq:gab1}
  g_{ab}(E) = \exp\left(-\sum_{n=1}^\infty R_{ab}^{(n)}(\Delta)\right),
\end{equation}
with the radiator components defined iteratively by
\begin{equation}
  \label{eq:Rabn}
  \begin{split}
  R_{ab}^{(n+1)}(\De) &= \int_0^{\De}d\De'\int_{\Cin}\!
  \frac{d^2\Om_k}{4\pi}\,w_{ab}(k)\>\prod_{i=0}^{n-1}U^{(i)}_{abk}(\De')\>
  \left[1-U^{(n)}_{abk}(\De')\right]\>,\\
  U^{(i)}_{abk}(\De) &=
  e^{-R_{ak}^{(i)}(\De)-R_{kb}^{(i)}(\De)+R_{ab}^{(i)}(\De)}\>.
  \end{split}
\end{equation}
We see immediately that at small $\De$, $R_{ab}^{(n)}(\De)$ is of
order $\De^{n+1}$ so that all terms contribute to SL accuracy. This is
the crucial difference with the case for global observables in which
the soft secondary branching contributes beyond SL level, see
discussion in subsection \ref{sec:global}.

In Appendix \ref{App:Iterative} we compute the first term for the
$\ee$ physical distribution $R_{p\bar p}^{(1)}(\De)$.  At small $\De$
we find (see \cite{DS})
\begin{equation}
  \label{eq:R1small}
  R_{p\bar{p}}^{(1)}(\Delta) = \Delta^2\left(\frac{\pi^2}{12}-
  \frac{\mbox{Li}_2(\tan^4\frac{\thin}{2})}{2}\right)
  +\cO{\Delta^3}\>,
\end{equation}
while at large $\Delta$ it becomes
\begin{equation}
  \label{eq:R1large}
  R_{p\bar{p}}^{(1)}(\Delta) = \Delta\ln\left(\Delta e^{\gamma_E-1}(1
  -\tan^4{\textstyle\frac{\thin}{2}})\right)+\cO{1}\>.
\end{equation}
As $\De$ increases, the higher contributions will also become
important.  However it is interesting to note the dependence on
$\thin$ -- even for quite large $\thin$ (of order 1) the distribution
$R_{p\bar{p}}^{(1)}$ is relatively insensitive to it, and in order to
get a noticeable $\thin$-dependence one has to move close to the limit
$\thin=\frac{\pi}{2}$.

\subsection{Asymptotic behaviour at large $\De$}
The most physically interesting study is of the behaviour at large
$\De$, which is discussed at length in section \ref{sec:Large}.
Here we summarise the method and important results.

To a first and quite crude approximation, we may treat $G_{ab}(\De)$
at large $\De$ as being approximately 1 around $a\!=\!b$, and very
small elsewhere, in other words having a peak of height 1 at
$a\!=\!b$. This cuts off the integral in equation \eqref{eq:basic2}
for the secondary emission $k$ near the emitters $a$ or $b$.
Physically this means that real gluons are only emitted close to one
of the emitters; in the other regions only virtual gluons contribute.
This is the same effect as described in \cite{DS}, where from
numerical results it was seen that secondary gluons were radiated in a
region around the primary partons, but that there also existed a large
empty ``buffer'' region extending to the edge of $\Cin$. Importantly,
this also means that the geometry of the $\Cin$ region becomes less
important, since what determines if a gluon is emitted or not in a
particular direction is no longer whether that direction is in $\Cin$
or $\Cout$, but if it is within the peak around one of the primary
emitters or not. Thus we find that the leading large-$\De$ behaviour
is independent of geometry. (Of course subleading terms are still
geometry-dependent.)

Of particular physical importance is the width of the distribution
peak, which is a measure of the size of the buffer.  We quantify this
in terms of a critical angle $\thc(\De)$ around parton $a$. Then we
introduce a function $h_\De(z)$ which for large $\De$ we take to give
the shape of the peak with normalised width. We derive coupled
evolution equations for $\thc(\De)$ and $h_\De(z)$ which we evolve
numerically to give what we assume to be stable limits for
$d\ln\thc/d\De$ and $h_\infty(z)$. At large $\De$ we then find
\begin{equation}
  \label{eq:thc}
  \thc(\De) \simeq \lam_a(\Cin)\cdot e^{-\frac{c}{2}\De}\>,
\end{equation}
where $c\approx 2.5$ is a universal constant. The subleading
behaviour, and in particular the normalization constant $\lam_a$, is
dependent on the geometry of $\Cin$ and on the position of parton $a$,
see section \ref{sec:Large}.  Thus at large $\De$ the region within
$\thc(\De)$ rapidly shrinks and the interjet region $\Cin$ becomes
more and more empty. This compares well with \cite{DS}, who suggest
from the numerical analysis the same functional form, with a value for
$c$ of about 3 (which is probably consistent within errors).

Then using the asymptotic peak shape $h_\infty(z)$ we obtain the
large-$\De$ behaviour for the whole distribution. Consider the physical
distribution for a general hard process given in terms of $G_{p_i
p_j}(\De)$ with $p_i$ and $p_j$ the hard primary jet directions which
in general form an angle $\theta_{ij}$. We derive the following
Gaussian behaviour at large $\De$
\begin{equation}
  \label{eq:gauss}
  G_{p_ip_j}(\De) \simeq e^{-\frac{c}{2}\De^2}
  \left(\frac{\lam_i\lam_j\,e^{-c'}}{1-\cos\theta_{ij}}\right)^{\De}
  \cdot f_{ij}\>,
\end{equation}
Here $c$ is the same universal constant in \eqref{eq:thc} and $c'$ a
second universal constant. The functions $f_{ij}$ and $\lam_i,\lam_j$
depend on the geometry $\Cin$ and the jet directions.  They arise as
integration constants in the large $\De$ approximation of the
evolution equation.

When the excluded jet region $\Cin$ is a pair of small cones of angle
$\thin$ centred on the jet directions $p_ip_j$ the width of the peak
becomes proportional to $\thin$
\begin{equation}
  \label{eq:thc0}
  \lam_i(\Cin)=\lam_j(\Cin)=\thin\cdot\hat\lam_0\>.
\end{equation}
We obtain
\begin{equation}
  \label{eq:smallthin}
  G_{p_i p_j}(\De) \simeq \left(\frac{\thin^2}
  {2(1-\cos\theta_{ij})}\right)^{\De}\cdot g(\De)\>,
  \qquad g(\De)\simeq e^{-\frac{c}{2}\De^2}
  \left(2\hat\lam_0^2\,e^{-c'}\right)^{\De}\hat f_0^2\>.
\end{equation}
At large $\De$ the Gaussian behaviour coming from the soft secondary
emission dominates over the bremsstrahlung contribution, which is the
first factor. For small $\thin$ the soft secondary branching
distribution $g(\De)$ becomes fully $\thin$-independent and
direction-independent.  The constants $\hat f_0,\hat\lam_0$ arise one
factor from each jet, since for small $\Cin$ the two jet contributions
factorize, and by symmetry they are equal. Thus in any hard QCD
process the distribution of energy emitted into the region excluding a
narrow cone around each hard jet is determined for each colour flow by
the usual bremsstrahlung radiator multiplied once for each emitting
dipole by the universal Gaussian function $g(\De)$.

If the weak dependence of $R_{p\bar{p}}^{(1)}(\De)$ on $\thin$ found
in \eqref{eq:R1large} is at all indicative of the general behaviour,
we may hope that the small cone approximation \eqref{eq:smallthin} for
$g(\De)$ could be valid even for quite large cones: this was indeed
seen in \cite{DS}.  This smooth behaviour is explained by the fact
that for small $\thin$ the critical region $\thc(\De)$ inside which
the branching develops scales proportionally to $\thin$, see
\eqref{eq:thc0}.

\subsection{Numerical results}
We try a numerical solution of \eqref{eq:basic3} to give the
quantitative behaviour of the secondary branching distribution
$g_{ab}(\De)$ in $\ee$ annihilation.  We set a grid of 40 bins in
$\phi$ and 80 bins in $\cos\theta$ and evolve \eqref{eq:basic3}
numerically.  Real-virtual cancellation is implemented by neglecting
the contribution with $k=a$ or $k=b$ in \eqref{eq:basic3}.  This may
be a crude approximation, due to the sharp behaviour of the function
$g_{ab}$ for $a$ near $b$, and needs improvement in order to have
complete control on the errors.  For what concerns the $\Delta$
behaviour of the solution, we find substantial agreement with the
results obtained in \cite{DS}.

As stated in the previous subsection, it is crucial to understand the
behaviour of the function $g_{ab}$ when we vary the opening angle
between $a$ and $b$. In figure \ref{fig:gab_th} we plot $g_{ab}$ as a
function of $1\!-\!\cos\theta_{ab}$ for three different values of
$\De$. We choose to fix $a$ along the thrust axis and let
$\theta_{ab}$ run from $0$ to $\thin$.  This distribution starts from
$1$ and then steeply decreases with increasing $\theta_{ab}$.  This
behaviour confirms the presence of a peak for the $g_{ab}$
distribution near $a=b$ which shrinks with increasing $\De$.  The form
of this distribution for large $\De$ will be discussed in detail in
the next section. From this figure we see that the asymptotic
behaviour is already settled at $\De=3$ where the distribution is
negligible away from the peak.

\EPSFIGURE[ht]{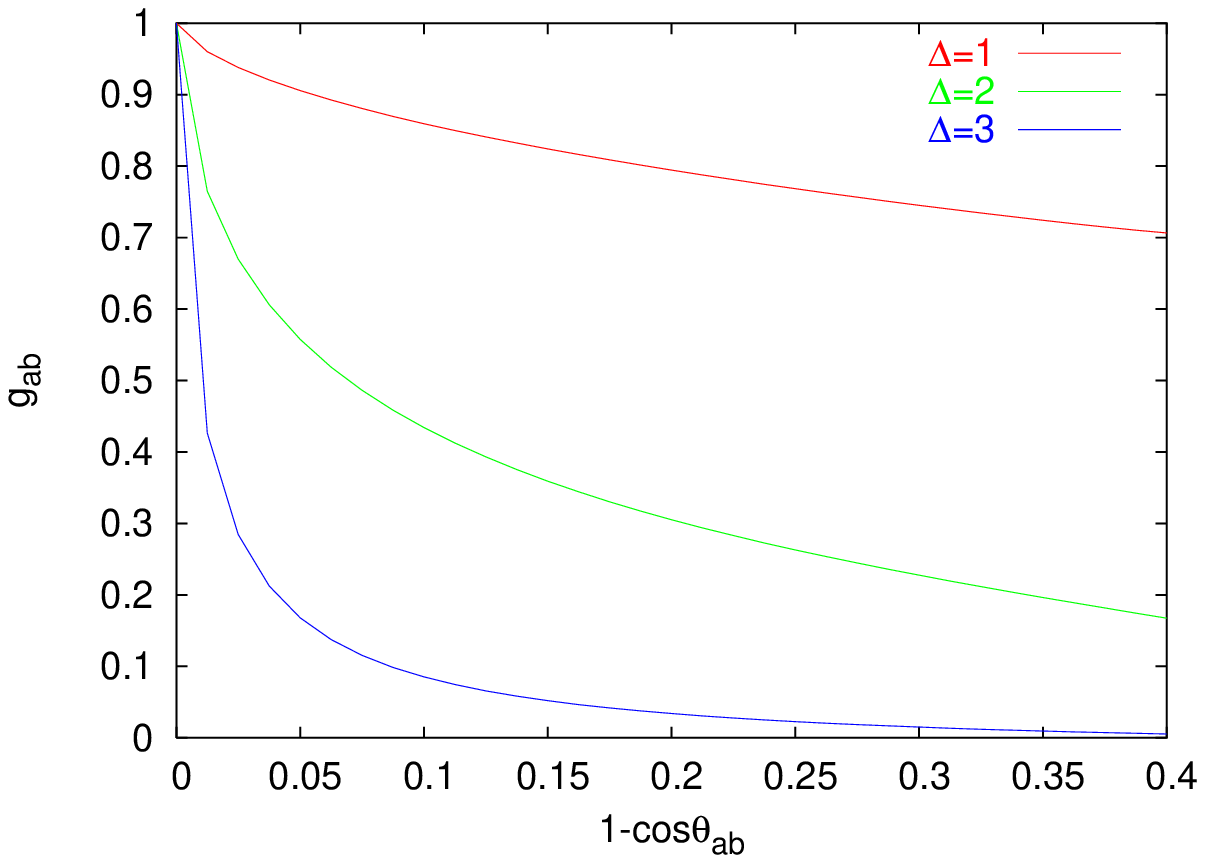,width=0.6\textwidth} {The distribution
  $g_{ab}$ for $\cos\thin=\half$ and $\theta_a=0$ as a function of
  $1\!-\!\cos\theta_{ab}$ for three different values of $\Delta$.
  \label{fig:gab_th}}

\section{Study of the large-$\De$ behaviour \label{sec:Large}}
We focus on $\ee$ annihilation with $\Cin$ given by a cone of size
$\thin$ around the two jets along thrust axis, see
\eqref{eq:region-ee}. We then generalize this analysis to the
distribution $G_{p_ip_j}$ required for other hard processes,
in which the two jets are not back-to-back. To obtain $G_{ab}(\De)$ in
the large $\De$ limit we need first to study its behaviour near the
peak at $a\!=\!b$. Using this information we can obtain the asymptotic
behaviour of $G_{ab}$ for any $a,b$.  These two stages are described
in the next two subsections. In a final subsection we consider the
particular case of $\Cin$ being a pair of small cones.

\subsection{Shape of the peak at $a\!=\!b$ \label{sec:peak}}
For each point $a$ inside the jet region $\Cin$, we can measure
the width of the peak in $b$ near $a$ by using
\begin{equation}
  \label{eq:thcrit}
  \int_{\Cin}\frac{d^2\Om_b}{4\pi}\,G_{ab}(\De) 
  \equiv 1-\cos\thc(\De)
  \approx \half\thc^2(\De)\>.
\end{equation}
In general this quantity depends on the geometry of the $\Cin$ region
and on the point $a$ chosen in $\Cin$. Here $\thc(\De)$ measures the
angle between $a$ and $b$ above which the distribution $G_{ab}$ is
suppressed. It is therefore a small angle which decreases as $\De$
increases. We have the initial value $\thc(\De\!=\!0)\!=\!\thin$.

Since the evolution of the peak is determined only by the distribution
around the peak, and not by points remote in phase-space, we have that
the shape of the peak in $G_{ab}(\De)$ depends only on the angle
$\theta_{ab}$ between $a$ and $b$. To measure this shape, suppose that
at large $\De$ and in the region near $a=b$ the distribution behaves
as
\begin{equation}
  \label{eq:hdef}
G_{ab}(\Delta) = 
h_\De(z)\>,\qquad z\>=\>\frac{\theta_{ab}^2}{2\thc^2(\De)}\>,
\end{equation}
for some function $h_\De(z)$ that from \eqref{eq:thcrit} for all large
$\De$ must satisfy
\begin{equation}
  \label{eq:hnorm}
  \int_0^\infty dz\,h_\De(z) = 1\>.
\end{equation}
The evolution equation \eqref{eq:basic2} simplifies in that the source
term involving $r_{ab}$ is negligible and the integration over solid
angle becomes an integration over a flat plane:
\begin{equation}
  \label{eq:hevolution}
  \begin{split}
    &\partial_\De h_\De(z) - 
    \frac{d\ln\theta_{\rm crit}^2}{d\De}\cdot zh_\De'(z) = 
    \frac{1}{2\pi}\int\frac{dx\,dy}{r_1^2 r_2^2}
    \left[h_\De(r_1^2 z)h_\De(r_2^2 z)-h_\De(z)\right]\>,
  \\
    &r_1^2 = x^2+y^2\>,\qquad r_2^2=(1-x)^2+y^2\>,\qquad
    \theta_{ak}=r_1\theta_{ab}\>,\qquad\theta_{kb}=r_2\theta_{ab}\>.
  \end{split}
\end{equation}
Integrating this equation over $z$ yields
\begin{equation}
  \label{eq:thcritevolution}
  \frac{d\ln\theta_{\rm crit}^2}{d\De} = \frac{1}{2\pi}
  \int_0^\infty dz\int\frac{dx\,dy}{r_1^2 r_2^2}
  \left[h_\De(r_1^2 z)h_\De(r_2^2 z)-h_\De(z)\right].
\end{equation}
Equations \eqref{eq:hnorm}-\eqref{eq:thcritevolution} form a coupled
system that can be evolved numerically from any suitable starting
function $h_0(z)$. Here $h_\De(z)$ is a function of only one variable
$z$ while the full distribution $G_{ab}(\De)$ is a function of three
angles, so the implementation is somewhat easier. 

There are many possible choices of starting function $h_0(z)$: it
need only be bounded between 0 and 1, with $h_0(0)=1$, normalised,
and sufficiently smooth. We used a selection of such functions, namely
\begin{equation}
h_0(z) = e^{-\left[z\Gamma(1+\frac{1}{n})\right]^n}\>, 
\>(n=\half,1,2,3,4)\>,\qquad
h_0(z) = \frac{2-z}{2}\Theta(2-z)\>,
\end{equation}
all of which showed the same asymptotic behaviour.
The evolution \eqref{eq:hevolution} settles down to a shape $h_\infty(z)$
shown in figure \ref{fig:hinfty}, which satisfies the equation 
\begin{equation}
  \label{eq:hinfty}
  \begin{split}
    &c\cdot zh_\infty'(z) = 
    \frac{1}{2\pi}\int\frac{dx\,dy}{r_1^2 r_2^2}
    \left[h_\infty(r_1^2 z)h_\infty(r_2^2 z)-h_\infty(z)\right]\>,
  \\
    &c\equiv -\frac{d\ln\theta_{\rm crit}^2}{d\De} = -\frac{1}{2\pi}
    \int_0^\infty dz\int\frac{dx\,dy}{r_1^2 r_2^2}
    \left[h_{\infty}(r_1^2 z)h_{\infty}(r_2^2 z)-h_{\infty}(z)\right]
    \approx 2.5\>,
  \end{split}
\end{equation}
with $c$ evaluated with an accuracy of about 10\%. This error is due
to limitations in the numerical analysis from the finite number of
$z$-points, and the numerical integrals performed at each stage, such
that evolving using the same starting function but with a different
number of points or precision in the integrals converges to a slightly
different numerical value for $c$. The step size in $\De$ needs to be
small enough to prevent instabilities developing before convergence is
seen --- we found the value 0.02 to be sufficiently small. The stated
uncertainty on the value of $c$ is a generous estimate. This error
could be improved with a more refined analysis.

We then conclude that $\thc(\De)$ has the behaviour at large $\De$
given in \eqref{eq:thc} with $\lambda_a$ an integration constant which
depends on $\thin$ and on the chosen point $a$.  The fact that
$h_{\infty}(z)$ is finite for any $z$ implies that at large $\De$ the
distribution $G_{ab}(\De)$ for $\theta_{ab}\ll1$ depends on
$\De,\thin$ and the point $a$ only through the function $\thc(\De)$.
The function $h_{\infty}(z)$ as well as the constant $c$ is universal.

\EPSFIGURE[ht]{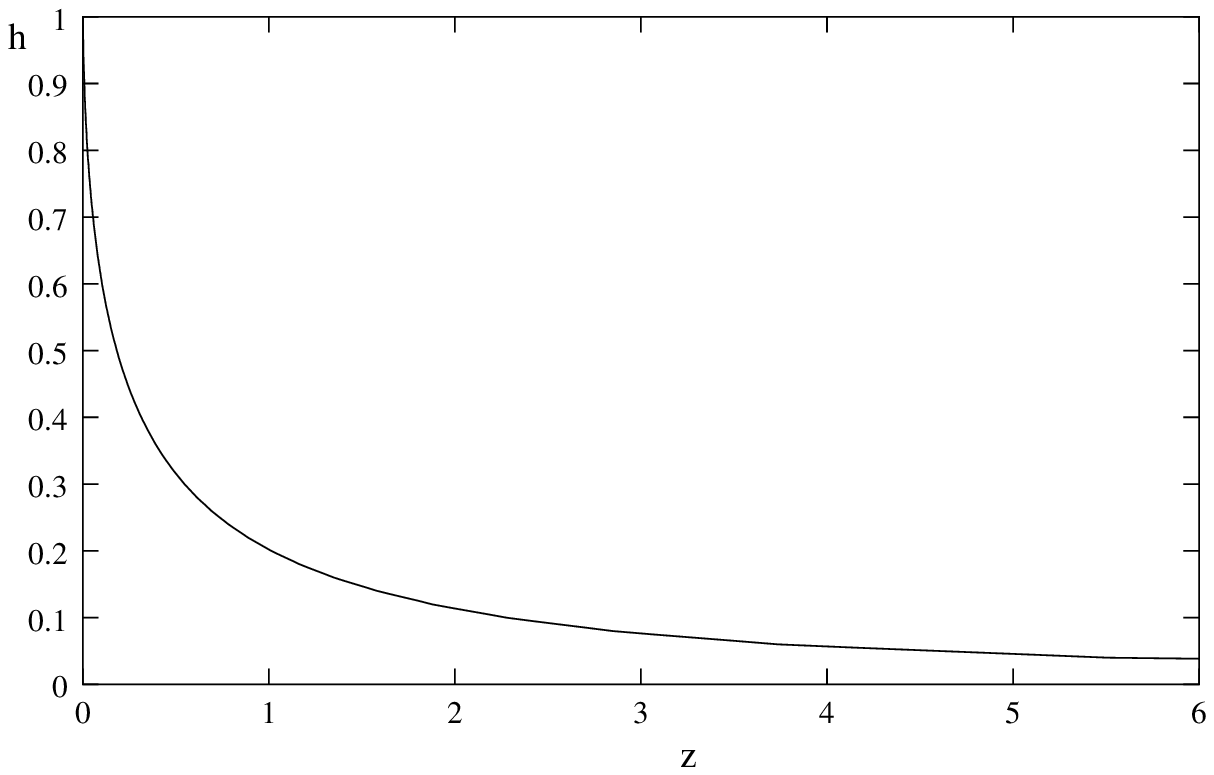,width=0.8\textwidth} 
{The large $\De$ peak, $h_\infty(z)$.\label{fig:hinfty}}

From the tail of the function $h_{\infty}(z)$ at large $z$ we can
estimate the large-$\De$ behaviour of $G_{ab}(\De)$ away from the peak
in the region
\begin{equation}
  \label{eq:tail}
  \thc(\De)\ll\theta_{ab}\ll1\>.
\end{equation}
To obtain this we rewrite equation \eqref{eq:hinfty}
in the form
\begin{equation}
  \label{eq:n1}
  c\cdot\frac{d\ln h_\infty}{d\ln z} = 
  \frac{1}{\pi}\int_{r_1<r_2}\frac{dx\,dy}{r_1^2 r_2^2}
  \left[h_\infty(r_1^2 z)-1+h_\infty(r_1^2 z)\left(
  \frac{h_\infty(r_2^2 z)}{h_\infty(z)}-1\right)\right]\>.
\end{equation}
As $z\to\infty$ the final term in the brackets vanishes and we obtain
\begin{equation}
  \label{eq:n2}
  c\cdot\frac{d\ln h_\infty}{d\ln z} = 
  \int_0^\infty\frac{dz'}{z'}\left[h_\infty(z')-1\right]\,
  \xi\left(\frac{z'}{z}\right)\>,
\end{equation}
with
\begin{equation}
\begin{split}
  \xi(x)=\left\{
  \begin{split}
  &\frac{1}{1-x}\>,\qquad\qquad\qquad\qquad\qquad\qquad\qquad\quad
  x<\frac{1}{4}\>, \\
  &\frac{2}{\pi(1-x)}\tan^{-1}\left(\frac{1-\sqrt{x}}{1+\sqrt{x}}
  \sqrt{\frac{2\sqrt{x}+1}{2\sqrt{x}-1}}\right)\>,
  \quad x>\frac{1}{4}\>.
  \end{split}\right.
\end{split}
\end{equation}
In the large $z$ limit the evolution equation becomes
\begin{equation}
  \label{eq:c'}
  c\cdot\frac{d\ln h_\infty}{d\ln z} \simeq -\ln z-c'\>,
  \qquad
  c' = \int_0^{1}\frac{dz'}{z'}\left[1-h_\infty(z')\right]
  - \int_{1}^\infty\frac{dz'}{z'}h_\infty(z')\>.
\end{equation}
Therefore, using \eqref{eq:thc} and \eqref{eq:hdef}, we conclude that
for $a$ and $b$ in the region \eqref{eq:tail} we have, in the large
$\De$ limit,
\begin{equation}
  \label{eq:smallthetaab}
  \ln G_{ab}(\De) \simeq -\frac{c}{2}\De^2-
  \left(c'+\ln\frac{\theta_{ab}^2}{2\lambda_a^2}\right)\De-
  \left(\frac{c''}{c}+\frac{c'}{c}\ln\frac{\theta_{ab}^2}{2\lambda_a^2}+
  \frac{1}{2c}\ln^2\frac{\theta_{ab}^2}{2\lambda_a^2}\right)\>.
\end{equation} 
The constants $c'$ and $c''$ can be determined by the function
$h_\infty(z)$ shown in figure \ref{fig:hinfty}.  This behaviour is valid
provided $a,b$ are away from the boundary of $\Cin$.

In taking the large-$z$ limit, we have neglected terms that contribute
at finite $z$ in moving from \eqref{eq:n1} to \eqref{eq:n2} and from
\eqref{eq:n2} to \eqref{eq:c'}. The leading correction is
$\cO{z^{-1}}$, and arises from the expansion of the $\xi$ function in
\eqref{eq:n2}. Thus the right hand side of \eqref{eq:smallthetaab} has
a correction $\cO{e^{-c\De}}$.

\subsection{The distribution off the peak \label{sec:oppositejets}}
Using the fact that we know the form of $G_{ab}(\De)$ near the peak at
$a\!=\!b$, we now determine the large-$\De$ behaviour of $G_{ab}(\De)$
for {\em any} $a$ and $b$ including the physical case in which $a$ and
$b$ are along the two hard jet directions ($\theta_{ab}\sim1$).  

We find that the geometry dependence in the integration regions of the
two terms in \eqref{eq:basic2} cancels for large $\De$.  To obtain
this result we write \eqref{eq:basic2} in the form
\begin{equation}
  \label{eq:basic6}
  \partial_\De\ln G_{ab} = 
  -\int_{\Cout}\frac{d^2\Om_k}{4\pi}\,w_{ab}(k)
  + \int_{\Cin}\frac{d^2\Om_k}{4\pi}
  w_{ab}(k)\left(\frac{G_{ak}\cdot G_{kb}}{G_{ab}}-1\right)\>.
\end{equation}
Since the first term in the second integral is negligible for $k$ not near
$a$ or $b$ we may write
\begin{equation}
  \label{eq:Gablog}
  \begin{split}
  \partial_\De\ln G_{ab} \simeq -\int_{\theta_{ak}^2,\theta_{kb}^2>\eps}
  &\frac{d^2\Om_k}{4\pi}w_{ab}(k)
  + \int_{\theta_{ak}^2<\eps}\frac{d^2\Om_k}{4\pi}
  w_{ab}(k)\left(\frac{G_{ak}\cdot G_{kb}}{G_{ab}}-1\right)\\
  &+\int_{\theta_{bk}^2<\eps}\frac{d^2\Om_k}{4\pi}
  w_{ab}(k)\left(\frac{G_{ak}\cdot G_{kb}}{G_{ab}}-1\right)\>,
  \end{split}
\end{equation}
where $\eps$ is a small parameter (larger than $\theta_{\rm crit}$) on
which the final answer should not depend. Here we have neglected the
ratio ${G_{ak}G_{kb}}/{G_{ab}}$ for $\theta_{ak},\,\theta_{bk}\gg\thc$
which vanishes at large $\De$.
Due to the cancellation of $\Cin$ and $\Cout$ in the integration
region, we have that, at large $\De$, the distribution $G_{ab}(\De)$
depends on the geometry and on the directions of $a$ and $b$ only
through ${\thc}_{,\,a}(\De)$ and ${\thc}_{,\,b}(\De)$, the critical
values in \eqref{eq:thc}.

We now show that the $\eps$-dependence in the various terms cancel.
The first term of \eqref{eq:Gablog} gives the contribution with a
cutoff around $a$ and $b$: it is, for small $\eps$,
\begin{equation}
  \int_{\theta_{ak}^2,\theta_{kb}^2>\eps}\frac{d^2\Om_k}{4\pi}w_{ab}(k)
  \simeq \ln\frac{2(1-\cos\theta_{ab})}{\eps}\>,
\end{equation}
while the second and third terms give the contributions from $k$ near
$a$ or $b$, which depend on the shape of the peak given above. We
write
\begin{equation}
  \int_{\theta_{ak}^2<\eps}\frac{d^2\Om_k}{4\pi}
  w_{ab}(k)\left(\frac{G_{ak}G_{kb}}{G_{ab}}-1\right)
  \simeq \frac{1}{2}\int_0^{\eps}\frac{d\theta_{ak}^2}{\theta_{ak}^2}
  \left(G_{ak}-1\right)
  \simeq -\frac{c'}{2}-\frac{1}{2}\ln\frac{\eps}{2{\thc^2}_{,\,a}}\>,
\end{equation}
with ${\thc}_{,\,a}$ the critical angle for the peak around $a$ and
$c'$ the constant given in \eqref{eq:c'}.  Here corrections vanish in
the large $\De$ limit. A similar equation holds for the integral
around $b$. Thus the $\eps$ dependence cancels and we obtain the
evolution equation at large $\De$:
\begin{equation}
  \label{eq:Gasympevol}
  \partial_\De\ln G_{ab} \simeq -c\De-c'
  -\ln\frac{1-\cos\theta_{ab}}{\lam_a\lam_b}\>.
\end{equation}
The asymptotic behaviour of $G_{ab}$ is therefore
\begin{equation}
  \label{eq:fine1}
  G_{ab}(\De) \simeq e^{-\frac{c}{2}\De^2}
  \left(\frac{\lam_a\lam_b\,e^{-c'}}{1-\cos\theta_{ab}}\right)^\De
  f_{ab}(\thin)\>,
\end{equation}
where $f_{ab}$ is independent of $\De$ at large $\De$.  For
$\theta_{ab}\ll1$ we have, see \eqref{eq:smallthetaab},
\begin{equation}
  \ln f_{ab} \simeq -\left(\frac{c''}{c}+\frac{c'}{c}
  \ln\frac{\theta_{ab}^2}{2\lam_a^2}+
  \frac{1}{2c}\ln^2\frac{\theta_{ab}^2}{2\lam_a^2}\right)\>.
\end{equation}
Because of this structure, we have that secondary branching is almost
collinear to the partons $a,b$ initiating the branching: in fact no
farther away than the width of the peak.  This implies that as long as
$a,b$ are not close to the boundary of $\Cin$ the geometry dependence
enters only through the parameter $\lam_a$ in $\theta_{\rm crit}$.

\subsection{Small $\Cin$ region}
Since the secondary branching is almost collinear to the hard primary
partons it becomes interesting to study the limit $\thin\to 0$.  In
particular we would like to see how the buffer region behaves when
the jet region is squeezed. We consider the two cases of $ab$ in
the same and in opposite jet regions.

\paragraph{Same jet region.}
We derive the evolution equation
\begin{equation}
  \label{eq:GRRsmall}
  \begin{split}
  &\partial_{\De}G_{ab}=-r_{ab}\cdot G_{ab}+
  \int_{|\vec{k}|<1}\frac{d^2\vec{k}}{2\pi}
  \frac{(\vec{a}-\vec{b})^2}{(\vec{a}-\vec{k})^2(\vec{k}-\vec{b})^2}\>
  \left[G_{ak}\,G_{kb}\,-\,G_{ab}\right]\,,\\
  &r_{ab} =\frac{1}{2}
  \ln\left(1+\frac{(\vec{a}-\vec{b})^2}{(1-a^2)(1-b^2)}\right)\>,\qquad
  \vec{a}=\frac{\theta_a}{\thin}\,\vec{n}_a\>,\qquad
  \vec{b}=\frac{\theta_b}{\thin}\,\vec{n}_b\>,
  \end{split}
\end{equation}
using rescaled angular variables $\vec{a},\vec{b}$. The contribution
to the evolution of $G_{ab}$ from the branching in the opposite region
vanishes for $\thin\to0$. This is an aspect of coherence of the QCD
radiation.  We notice that the explicit $\thin$ dependence has
disappeared from the equation. This implies that $\thc(\De)$ scales
with $\thin$ so that
\begin{equation}
  \label{eq:thc0a}
  \lam_a(\Cin)=\thin\cdot\hat\lam(\vec{a})\>,
\end{equation}
giving \eqref{eq:thc0} in the case $\vec{a}=0$. We have then that the
buffer region still expands by increasing $\De$.  By performing the
same analysis as in subsection \ref{sec:peak} we find, at large $\De$,
for $ab$ in the same jet region but away from the peak, the behaviour
\begin{equation}
  \label{eq:same}
  G_{ab}(\De) \simeq e^{-\frac{c}{2}\De^2}
  \left(\frac{2\hat\lam(\vec{a})\,\hat\lam(\vec{b})\,e^{-c'}}
  {(\vec{a}-\vec{b})^2}\right)^\De\hat f(\vec{a},\vec{b})\>.
\end{equation}

\paragraph{Opposite jet regions.}
We introduce now rescaled angular variables for the two back-to-back
jets (left and right jet). For $\thin\to0$ the radiator is
\begin{equation}
  \label{eq:rabsum}
  r_{ab}=-2\ln\frac{\thin}{2}
  -\frac{1}{2}\ln(1\!-\!a^2)-\frac{1}{2}\ln(1\!-\!b^2)\>,\qquad
  \vec{a}=\frac{\theta_a}{\thin}\,\vec{n}_a\>,\quad
  \vec{b}=\frac{\pi-\theta_b}{\thin}\,\vec{n}_b\>.
\end{equation}
This explicit $\thin$ dependence implies that $G_{ab}(\De)$ is given
by $\thin^{2\De}$ (the bremsstrahlung contribution) times a function
of the rescaled variables $\vec{a},\vec{b}$, whereas the same jet
distribution depends only on the rescaled quantities.

The $a$ and $b$ contributions enter \eqref{eq:rabsum} independently.
This is due to the fact that $w_{ab}(k)$ splits into a sum of right
and left pieces.  As a consequence the evolution in the right and left
regions develops independently and $G_{ab}(\De)$ factorizes
\begin{equation}
  \label{eq:GRLfact}
  G_{ab}(\De)\simeq \left(\frac{\thin}{2}\right)^{2\De}
  \cdot F^{\rm R}_a(\De)\cdot F^{\rm L}_b(\De)\>.
\end{equation}
Here $F^{\rm R}_a$ and $F^{\rm L}_b$ satisfy the right and left
evolutions and for the first we find
\begin{equation}
  \label{eq:GRLsmall}
\partial_{\De}F_{a}^{\rm R} = 
\frac{1}{2}\ln(1-a^2)\cdot F_{a}^{\rm R}+
\int\frac{d^2\vec{k}}{2\pi\,(\vec{a}-\vec{k})^2}\>
\left[G_{ak}^{\RR}\,F_{k}^{\rm R}\,-\,F_{a}^{\rm R}\right]\,,
\end{equation}
where $G_{ak}^{\RR}$ is the distribution for $a$ and $k$ in the same
(right) jet region discussed above.  Proceeding as in subsection
\ref{sec:oppositejets} we find the following asymptotic behaviour for
the distribution with $ab$ in opposite jet regions
\begin{equation}
  \label{eq:GabRLfine}
  G_{ab}(\De)\simeq e^{-\frac{c}{2}\De^2}
  \left(\frac{\thin^2\,\hat\lam(\vec{a})\,\hat\lam(\vec{b})\,e^{-c'}}
  {2}\right)^{\De}\>\hat f(\vec{a})\hat f(\vec{b})\>,
\end{equation}
with $c'$ the integration constant given in \eqref{eq:c'} and $\hat f$
an integration constant depending only on the rescaled variable. This
expression shows that for small $\thin$ the function $f_{ab}(\thin)$
in \eqref{eq:fine1} factorizes and depends on the rescaled
variables. The $\thin$-dependence is the one given by the
bremsstrahlung contribution. The secondary branching contribution does
not depend on $\thin$ and is decreasing with $\De$ with a Gaussian
behaviour. For the physical $\ee$ distribution we set $a^2=b^2=0$.

A generalization of this result is to the case where the two jets
$p_ip_j$ are not back-to-back but are set at an angle $\theta_{ij}$.
The asymptotic behaviour is given by the expression \eqref{eq:GabRLfine}
with $\{2\}^{-\De}$ replaced by $\{1-\cos\theta_{ij}\}^{-\De}$
and with $\theta_a$ the angle of $a$ with respect to the jet $p_i$ and
$\theta_b$ the angle of $b$ with $p_i$. The physical distribution
$G_{p_ip_j}(\De)$ is again obtained by setting $a^2\!=\!b^2\!=\!0$.

\section{Discussion \label{sec:disc}}
We summarize here the essential physical points of our results on
energy flow away from hard jets.  The interjet distribution reduces to
$G_{p\bar p}(\De)$ in \eqref{eq:Sig-ee1} for $\ee$, while for DIS and
hadron-hadron collisions it is given in terms of $G_{p_ip_j}(\De)$,
see \eqref{eq:Sig-DIS2} and \eqref{eq:etaptEout}.  These distributions
depend on the geometry of $\Cout$ and on the jet directions.

A characteristic of interjet observables is that the leading
contribution to their distributions are SL, originating from soft
emission at large angles. Collinear singularities here are subleading.
The contribution from bremsstrahlung emission from the primary hard
partons does not give the full SL structure. Indeed we find (see
\eqref{eq:gab})
\begin{equation}
  \label{eq:Gij}
  G_{p_ip_j}(\De)=e^{-R^{(0)}_{p_ip_j}(\De)}\cdot g_{p_ip_j}(\De)\>,
\end{equation}
with $R^{(0)}_{p_ip_j}(\De)$ the SL Sudakov radiator for
emission into $\Cout$. The additional SL factor $g_{p_ip_j}$ is
generated by soft secondary emission, which can be described by
successive colour singlet soft dipole emission. This implies that, in
the present formulation, we work in the large $N_c$ approximation.
The soft secondary branching develops with decreasing energy inside
the unobserved jet region $\Cin$ until a soft gluon enters $\Cout$.
The emission is also effectively angular ordered.

At large $\De$ the distribution $G_{p_i p_j}(\De)$ decreases with a
universal Gaussian behaviour, see equation \eqref{eq:gauss}, with a
process-independent coefficient $c$ that we have estimated to be
$c\approx 2.5$ with a $10\%$ accuracy. Since the bremsstrahlung
radiator is proportional to $\De$ we see that the large $\De$
behaviour is dominated by the soft secondary branching contribution
$g_{p_ip_j}(\De)$.

The origin of the universal Gaussian behaviour in $\De$ is a
consequence of the structure of the branching. The secondary branching
generating $g_{p_ip_j}(\De)$ develops within a small cone $\thc(\De)$
around the hard primary partons $p_i$ and $p_j$, which decreases
exponentially for large $\De$ according to \eqref{eq:thc}, again
governed by the universal coefficient $c$. It is the development of
soft secondary branching in this peak region which generates the
Gaussian behaviour in \eqref{eq:gauss}. At large $\De$ the region
within $\thc(\De)$ shrinks generating an empty buffer region first
observed in \cite{DS}. For the case of figure \ref{fig:gab_th} we have
estimated that the asymptotic regime sets in around $\De=3$. At this
value the distribution is negligible away from the peak and approaches
the asymptotic limit of figure \ref{fig:hinfty}.

Taking the jet region $\Cin$ to be an arrangement of small cones
centred on each hard jet leads to a universal secondary emission
function $g(\De)$ that depends neither on the size of the cones nor on
the directions of the jets (see \eqref{eq:smallthin}).

The distribution $G_{p_i p_j}(\De)$ is a function of $E,\Eout$ and the
coupling $\as$ through the quantity $\De$ defined in \eqref{eq:rad}.
This involves an integral over the running coupling at low scales, and
so acquires non-perturbative power corrections. These corrections can
be estimated by the same method \cite{NPstandard} used for the
standard shape variables in $\ee$ and are expressed in term of a
unique parameter $\cp$ given by the integral over the running coupling
in the low energy region.  We obtain
\begin{equation}
   \label{eq:DNP}
   \De \equiv
   \int_{0}^{E}\frac{d\om}{\om}\bas(\om)\,[1-e^{-\om/\Eout}]
   \simeq \int_{\Eout}^{E}\frac{d\om}{\om}\bas(\om)+N_c \frac{\cp}{2\Eout}\>.
\end{equation}
The non-perturbative parameter $\cp$ (for the normalization used here
see \cite{cp}) is measured both at LEP~\cite{Exp-shape} and at
HERA~\cite{HERA}.

In hadron-hadron collisions one expects \cite{MW,FHT,MC} additional
large contributions to interjet observables coming from a soft
underlying event generated by the incoming hadrons. We can write
\begin{equation}
    \label{eq:underlying}
    \Eout= \sum_{i\in\Cout}\om_i +\Eout^{\rm soft}\>,
\end{equation}
where the sum includes all partons produced in the hard process, and
the quantity $\Eout^{\rm soft}$ is the contribution of the soft
underlying event.  Assuming the underlying emission and hard emission
factorize, the final answer will be given by the distribution studied
here with the replacement
\begin{equation}
  \De(E,\Eout)\to \De(E,\Eout-\Eout^{\rm soft}) \>.
\end{equation}
Up to now in hadron-hadron collisions only the mean value of $\Eout$
has been investigated both theoretically and experimentally.  One
shows that the hard contribution, which can be reliably computed by
fixed order QCD calculation\footnote{The PT calculation in \cite{MW}
  has been done only to $\as^3$ order, now it is possible to perform
  the calculation at order $\as^4$ \cite{3jetNLO} so that one can
  control the QCD scales.}, does not fit the data \cite{MW,UA1,FHT}
but requires a sizable $\Eout^{\rm soft}$ contribution.

The resummed PT distribution here discussed provides a way to further
study the underlying event, in particular its factorization properties
and its size.

\section*{Acknowledgements}
We would like to thank Gavin Salam, Mrinal Dasgupta and Giulia
Zanderighi for helpful discussions and suggestions.

\newpage
\appendix
\section{Interjet observables in DIS and hadron-hadron collisions}
\label{App:DIShh}
We introduce first the physical observables and then the QCD
resummation.

\subsection{The observable} 
An example of an interjet distribution in DIS is given by
\begin{equation}
  \label{eq:Sig-DIS}
  \Sigma_{\rm DIS}(\Eout)=
  \left\{\sum_n\int\frac{d\sigma_n}{d\xB\,dQ^2}\right\}^{-1}
  \sum_n\int\frac{d\sigma_n}{d\xB\,dQ^2} \cdot 
  \Theta\!\left(\Qout-\sum_{h\,\in\,\Cout}\om_h\right),
\end{equation}
with $d\sigma_n/d\xB dQ^2$ the $n$-hadron distribution for fixed
Bjorken variable $\xB$ and $Q^2$. The events are dominated by the
incoming and the outgoing hard jet. In the Breit frame the situation
is similar to $\ee$ with both jets aligned along the beam direction.
We may define the $\Cout$ region by
\begin{equation}
  \label{eq:regionDIS}
  -\eta^{(-)}_{\rm in}\> < 
  \>\ln\tan \frac{\theta_h}{2}\> <
  \>\eta^{(+)}_{\rm in}\>.
\end{equation}
An example of an interjet distribution in hadron-hadron collisions 
with hard jets is
\begin{equation}
  \label{eq:Sig-hh}
  \Sigma_{\rm hh}(\Eout)=
  \left\{\sum_n\int\frac{d\sigma_n}{d\eta\,dP_t} \right\}^{-1}
  \sum_n\int\frac{d\sigma_n}{d\eta\,dP_t} \cdot 
  \Theta\!\left(\Qout-\sum_{h\,\in\,\Cout}\om_h\right),
\end{equation}
with $d\sigma_n/d\eta dP_t$ the $n$-hadron distribution associated to
the emission of a jet with rapidity $\eta$ and transverse momentum
$P_t$.  The jet could be defined by a $k_t$-jet finding algorithm
\cite{ktjet}.  Here the event is dominated by the presence of an
additional recoiling outgoing jet. The interjet region $\Cout$ should
be defined by avoiding the two incoming and the two outgoing hard
jets.  It is given for instance by fixing a region in the rapidity
$\eta_h$ and azimuthal angle $\phi_h$ of emitted hadrons with respect
to the triggered jet, see \cite{UA1,FHT} .

The QCD resummation discussed for $\ee$ annihilation in section
\ref{sec:QCD} can be generalized to the above processes due to the
factorization of the parton process (see for instance
\cite{factorization}) as we shall discuss.

\subsection{QCD resummation: DIS case}
To express the result for $\Sigma_{\rm DIS}(\Eout)$ we write the DIS
cross section in the form
\begin{equation}
  \label{eq:Sig-DIS1}
  \frac{d\sigma}{d\xB dQ^2}=\int dx_1\,
  \frac{d\sigma_{\gam+p_1\to p_2}(x_1,\mu)}{dx_1d\xB dQ^2}\>,
  \quad \mu\sim Q\>,
\end{equation}
where $d\sigma_{\gam+p_1\to p_2}(x_1,\mu)$ is the factorized parton
distribution including the parton density function at the hard scale
$\mu\sim Q$ and the squared matrix element of the hard vertex
$\gam^*+p_1\to p_2$. Here $x_1$ is the momentum fraction of the hard
parton $p_1$ coming into the hard vertex in the Breit frame.  In the
soft limit $p_1$ and $p_2$ correspond to the directions of the
incoming and outgoing jets respectively.

By performing the same analysis as done in section \ref{sec:QCD} we
obtain
\begin{equation}
  \label{eq:Sig-DIS2}
  \Sigma_{\rm DIS}(\Eout)=\left\{\frac{d\sigma}{d\xB dQ^2}\right\}^{-1}
  \!\!\!\int dx_1\,\frac{d\sigma_{\gam+p_1\to p_2}(x_1,\mu)}{dx_1d\xB dQ^2}
  \cdot G_{p_1p_2}(Q,\Eout)\>.
\end{equation}
The factorization scale remains at $\mu$ since no observation is made
in the collinear region.  Here $G_{p_1p_2}(Q,\Eout)$ is the
distribution introduced in \eqref{eq:GR-ee} with $p_1p_2$ aligned, in
the Breit frame, along the beam direction.

\subsection{QCD resummation: hadron-hadron case}
The situation here is more involved since we have four jets (two
outgoing and two incoming). 
There are various hard elementary
processes
\begin{equation}
  \label{eq:hproc}
  p_1+p_2\>\to\>p_3+p_4\>,
\end{equation}
according to the nature ($q,\bar q$ or $g$) of the four involved
partons.  
They are:
\begin{equation}
\label{eq:hproc-list}
\begin{split}
{\rm A}\qquad & \> q(p_1)+q'(p_2)\>\to\>q(p_3)+q'(p_4)\>,\\
{\rm B}\qquad & \> q(p_1)+q(p_2)\>\to\>q(p_3)+q(p_4)\>,\\
{\rm C}\qquad & \> q(p_1)+\bar q(p_2)\>\to\>g(p_3)+g(p_4)\>,\\
{\rm D}\qquad & \> g(p_1)+g(p_2)\>\to\>g(p_3)+g(p_4)\>,
\end{split}
\end{equation}
together with those obtained by crossing transformations.

Energy flow in interjet regions for this process has been studied in
\cite{BKS} as far as the SL bremsstrahlung radiator piece.  To
construct the SL contribution from the soft secondary branching we
need to represent the soft $n$-gluon emission in each of the hard
processes \eqref{eq:hproc} as a combination of colour singlet dipole
emissions \eqref{eq:M2n} from the four hard partons.  The square
amplitude for the emission in \eqref{eq:hproc} of a single soft gluon
$k$ has been written in \cite{EMW}.  Taking the large $N_c$ limit,
this result can be expressed \cite{MWMC} as a sum of dipole emissions
$w_{ij}(k)$ from pairs $p_ip_j$ of hard partons with the factorized
structure
\begin{equation}
  \label{eq:ffffg}
  M^2_{f_1f_2f_3f_4\,,k}\to M^2_{f_1f_2f_3f_4}\cdot  
  \sum_{\gam}C^{(\gam)}_{f_1f_2f_3f_4} \sum_{\VEV{ij}\,\in\,\gam}
  \frac{\bas}{\om_k^2}\,w_{ij}(k)\>,
  \qquad
  \sum_{\gam}C^{(\gam)}_{f_1f_2f_3f_4}=1\>,
\end{equation}
where $M^2_{f_1f_2f_3f_4}$ is the exact hard elementary $2\!\to\!2$
distribution for the process \eqref{eq:hproc} with $f_i$ the type of
parton $p_i$. The coefficients $C^{(\gam)}_{f_1f_2f_3f_4}$ are 
functions of the kinematical invariants $\hat{s}\!=\!(p_1\!+\!p_2)^2$ ,
$\hat{t}\!=\!(p_1\!-\!p_3)^2$, $\hat{u}\!=\!(p_1\!-\!p_4)^2$ and can be found
in~\cite{MWMC}.  Here $\gam$ represents the set of all
colour-connected partons in the specific hard process and the last sum
is over all colour connected pairs $ij$ in the set $\gam$.

The simplest process is $qq'\to qq'$ with only $\VEV{f_1f_4}$
and $\VEV{f_2f_3}$ colour connections. Here the coefficients are
independent of the invariants and, from symmetry, we have
\begin{equation}
  \label{eq:qqqq}
  \sum_{\gam}C^{(\gam)}_{qq'qq'} \sum_{\VEV{ij}\in \gam}w_{ij}(k)
  =w_{14}(k)+w_{23}(k)\>.
\end{equation}
The most complex process is $gg\to gg$ for which all possible colour
connections contribute.

The distribution for the emission in the hard process \eqref{eq:hproc}
of $n$ additional soft gluons in the strongly energy ordered region is
obtained by performing the analysis done in \cite{BCM,FMR} in the
large $N_c$ limit. Each additional soft gluon contributes to one of
the hard dipole emissions as done in \eqref{eq:M2n}. This allows us to
extend the analysis of section \ref{sec:QCD} and deduce the interjet
$\Eout$ distribution in hadron-hadron collisions with a hard jet of
rapidity $\eta$ and transverse momentum $P_t$.

Consider the hadronic cross section for this process which we write in
the form
\begin{equation}
  \label{eq:DY}
  \frac{d\sigma}{d\eta dP_t}=
  \sum_{\{f_i\}}\int dx_1dx_2 
  \frac{d\sigma_{f_1f_2\to f_3f_4}(x_1x_2,\mu)}{dx_1dx_2d\eta dP_t}\>,
  \quad \mu\sim P_t\>,
\end{equation}
where $d\sigma_{f_1f_2\to f_3f_4}$ is the parton process integrand
which includes parton density functions and hard parton distribution
$M^2_{f_1f_2f_3f_4}$ for the elementary process \eqref{eq:hproc}. The
normalized $\Eout$ distribution in a given interjet region $\Cout$ is
then given by
\begin{equation}
  \label{eq:etaptEout}
  \begin{split}  
  \Sigma_{\rm hh}(\Eout)
  &=\left\{\frac{d\sigma}{d\eta dP_t}\right\}^{-1} \!\!
  \sum_{\{f_i\}}\int dx_1dx_2 
  \frac{d\sigma_{f_1f_2\to f_3f_4}(x_1x_2,\mu)}{dx_1dx_2d\eta dP_t}
\Sigma_{f_1f_2f_3f_4}(P_t,\Eout)\>,\\
&\Sigma_{f_1f_2f_3f_4}(P_t,\Eout)=\sum_{\gam}C^{(\gam)}_{f_1f_2f_3f_4} 
 \prod_{\VEV{ij}\,\in\,\gam}G_{p_ip_j}(P_t,\Eout)\>,      
  \end{split}
\end{equation}
where $G_{p_ip_j}(P_t,\Eout)$ is the distribution introduced in
\eqref{eq:GR-ee} associated with the dipole $p_i,p_j$.
In the soft limit we can take $p_1p_2$ in
the direction of the two incoming hadrons and $p_3p_4$ in the
direction of the two outgoing hard jets.
We report the explicit form of $\Sigma_{f_1f_2f_3f_4}(P_t,\Eout)$ for
the elementary processes in \eqref{eq:hproc-list}:
\begin{equation}
\label{eq:Sigma-hproc}
\begin{split}
H^{\rm A}\,\Sigma^{\rm A}
&\!=\!h^{\rm A}(\hat{s},\hat{t},\hat{u})\,G_{14}G_{23}\>,\\
H^{\rm B}\,\Sigma^{\rm B}&\!=\! 
h^{\rm B}(\hat{s},\hat{t},\hat{u})\>G_{14}G_{23}
+h^{\rm B}(\hat{s},\hat{u},\hat{t})\>G_{13}G_{24}\,, \\
H^{\rm C}\,\Sigma^{\rm C}&\!=\!
h^{\rm C}(\hat{s},\hat{t},\hat{u})\>G_{34} G_{13}
G_{24}+h^{\rm C}(\hat{s},\hat{u},\hat{t})\>G_{34} G_{14}
G_{23}\,,\\
H^{\rm D}\,\Sigma^{\rm D}&\!=\!
h^{\rm D}(\hat{s},\hat{t},\hat{u})G_{12}G_{24}
G_{43}G_{31}+h^{\rm D}(\hat{s},\hat{u},\hat{t})\>G_{12}G_{23}
G_{34}G_{41}\\&
+h^{\rm D}(\hat{u},\hat{t},\hat{s})\>G_{14}G_{42}
G_{23}G_{31}\,,
\end{split}
\end{equation}
with $G_{ij}=G_{p_ip_j}(P_t,\Eout)$ and the functions $h^{({\rm
    A,B,C,D})}$ given in table~\ref{tab:hproc-fun}. Here
\begin{equation}
\begin{split}
&H^{\rm A}=h^{\rm A}(\hat{s},\hat{t},\hat{u})\,,\\
&H^{\rm B}=
h^{\rm B}(\hat{s},\hat{t},\hat{u})+h^{\rm B}(\hat{s},\hat{u},\hat{t})\,,\\
&H^{\rm C}=
h^{\rm C}(\hat{s},\hat{t},\hat{u})+h^{\rm C}(\hat{s},\hat{u},\hat{t})\,,\\
&H^{\rm D}=
h^{\rm D}(\hat{s},\hat{t},\hat{u})+h^{\rm D}(\hat{s},\hat{u},\hat{t})+
h^{\rm D}(\hat{u},\hat{t},\hat{s})\>,
\end{split}
\end{equation}
are the square amplitudes for the $2\to2$ elementary processes
\eqref{eq:hproc-list}.

\TABLE[t]{
\centerline{
\begin{tabular}{|c|}
\hline 
$h^{\rm A}(s,t,u)=
g^4\frac{C_F}{N_c}\left(\frac{s^2+u^2}{t^2}\right)$ \\
$h^{\rm B}(s,t,u)=h^{\rm  A}(s,t,u)+
2g^4\frac{C_F}{N_c^2}\frac{s}{t}$ \\
$h^{\rm C}(s,t,u)=g^4C_F\frac{u}{t}
\left(\frac{t^2+u^2}{s^2}-\frac{1}{N_c^2}\right)$\\
$h^{\rm D}(s,t,u)=2g^4\left(\frac{N_c^2}{N_c^2-1}\right)
\left(1-\frac{tu}{s^2}
-\frac{su}{t^2}+\frac{u^2}{st}\right)$\\
\hline
\end{tabular}}
\caption{The functions $h^{({\rm A,B,C,D})}$ taken from Ref. \cite{MWMC}.}
\label{tab:hproc-fun}
}

The structure in \eqref{eq:Sigma-hproc} in terms of the factorized
distributions $G_{ij}$ is derived from eq.~66 of Ref.~\cite{MWMC} in
which we have identified $2C_F$ with $C_A$ since we work in the large
$N_c$ limit. Going beyond the large $N_c$ approximation requires the
analysis of colour interference between different hard partons.  At
the moment this can be done only for the bremsstrahlung contribution,
elegantly computed in~\cite{BKS}, see also \cite{G&K}.

In conclusion, for all considered hard processes, the interjet
distributions are described by the universal function
$G_{p_ip_j}(E,\Eout)$ which depends on the two hard jet directions
$p_i$ and $p_j$ (incoming or outgoing) and on the interjet region
$\Cout$. 

\section{Theorem: $0\le G_{ab}(\De)\le1$}
\label{App:Proof}
In this appendix we show that the solution of the differential
equation \eqref{eq:basic2} with the given boundary condition is
bounded below and above by 0 and 1. Thus our distribution is
physically meaningful.

Proof:
\begin{enumerate}
\item $G_{ab}(\De)$ is a continuous function of $\De$ and of $a$ and $b$.
\item $G_{ab}(0)=1\>\forall\>a,b$.
\item If $a=b$ then $G_{ab}(\De)=1\>\forall\>\De$.
\item Suppose $\exists\>a',b'$ and $\De'$ with $G_{a'b'}(\De')>1$.
Then by continuity $\exists\>a'',b''$ and $\De''$ with 
$0\le\De''<\De'$, $G_{ab}(\De'')\le1\>\forall\>a,b$, and $a''\ne b''$,
$G_{a''b''}(\De'')=1$, $\partial_\De G_{a''b''}(\De'')\ge0$.
\item But by \eqref{eq:basic2} we have 
$\partial_\De G_{a''b''}(\De'')<0$ which gives a contradiction.
Therefore $G_{ab}(\De)\le1\>\forall\>a,b,\De$.
\item Suppose now $\exists\>a',b'$ and $\De'$ with $G_{a'b'}(\De')<0$.
Then by continuity $\exists\>a'',b''$ and $\De''$ with 
$0\le\De''<\De'$, $G_{ab}(\De'')\ge0\>\forall\>a,b$, and $a''\ne b''$,
$G_{a''b''}(\De'')=0$, $\partial_\De G_{a''b''}(\De'')\le0$.
\item But by \eqref{eq:basic2} we have 
$\partial_\De G_{a''b''}(\De'')>0$ which gives a contradiction.
Therefore $G_{ab}(\De)\ge0\>\forall\>a,b,\De$.
\end{enumerate}

\section{Iterative solution \label{App:Iterative}}
Here we calculate the first term $R_{p\bar{p}}^{(1)}(\De)$ in the
iterative solution described in section \ref{sec:iterative}. First
we have that for any $a,b$ the bremsstrahlung emission is given by
\begin{equation}
\begin{split}
r_{ab} &= \frac{s_a}{2}\ln\frac{\cos\theta_a+\cos\thin}
{\cos\theta_a-\cos\thin} + \frac{s_b}{2}\ln
\frac{\cos\theta_b+\cos\thin}{\cos\theta_b-\cos\thin}
+\frac{s_a+s_b}{4}\times\\
&\hspace{-0.5cm}
\ln\frac{(1\!+\!\cos^2\thin)(1\!+\!\cos\theta_a\cos\theta_b)\!-\!
2\cos\thin(\cos\theta_a\!+\!\cos\theta_b)\!-\!
\sin^2\thin\sin\theta_a\sin\theta_b\cos\phi_{ab}}
{(1\!+\!\cos^2\thin)(1\!+\!\cos\theta_a\cos\theta_b)\!+\!
2\cos\thin(\cos\theta_a\!+\!\cos\theta_b)\!-\!
\sin^2\thin\sin\theta_a\sin\theta_b\cos\phi_{ab}},
\end{split}
\end{equation}
where $s_k=1$ for $\cos\theta_k>0$ and $s_k=-1$ for $\cos\theta_k<0$.
Therefore we obtain $R_{ab}^{(0)}$ in the following cases
\begin{equation}
\begin{split}
&R_{p\bar{p}}^{(0)}(\Delta) = \Delta\cdot
\ln\frac{1+\cos\thin}{1-\cos\thin}\>,\\
&R_{pk}^{(0)}(\Delta) = \Delta\cdot\frac{s_k}{2}
\ln\frac{(1-\cos\thin)(\cos\theta_k+\cos\thin)}
{(1+\cos\thin)(\cos\theta_k-\cos\thin)}\>,\\
&R_{k\bar{p}}^{(0)}(\Delta) = \Delta\cdot\frac{s_k}{2}
\ln\frac{(1+\cos\thin)(\cos\theta_k+\cos\thin)}
{(1-\cos\thin)(\cos\theta_k-\cos\thin)}\>.
\end{split}
\end{equation}

Now using the definition \eqref{eq:Rabn} we explicitly calculate 
\begin{equation}
\label{eq:Rab1}
R_{p\bar p}^{(1)}(\De) = \int_0^{\De}d\De'\int_{\Cin}\!
\frac{d^2\Om_k}{4\pi}\,w_{p\bar p}(k)\>
\left[1-U^{(0)}_{p \bar pk}(\De')\right]\>,
\end{equation}
where
\begin{equation}
U^{(0)}_{p\bar p k}(\De)=
\left(\frac{|\cos\theta_k|-\cos\thin}
           {|\cos\theta_k|+\cos\thin}\right)^{\De}
\left(\frac{1+\cos\thin}
           {1-\cos\thin}\right)^{\De}\>.
\end{equation}
For large $\De$ the integrand $[1-U^{(0)}_{p\bar p k}(\De)]$ forces
$k$ to stay away from the thrust axis by an small angle of order
$\De^{-1}$. Changing integration variables we have
\begin{equation}
\label{eq:Rpp1}
R_{p\bar{p}}^{(1)}(\Delta) = \int_0^\Delta d\Delta'\int_0^1 dx
\left(\frac{1-x^{\Delta'}}{1-x}-
\frac{1-x^{\Delta'}}{\cot^4\frac{\thin}{2}-x}\right)\>.
\end{equation}
At small $\De$ this gives the result \eqref{eq:R1small}, while at large 
$\De$ we obtain \eqref{eq:R1large}.

\end{document}